\documentstyle[prl,aps,twocolumn]{revtex}

\begin{document}
\title{Coherent molecular solitons in Bose-Einstein condensates}
\author{P. D. Drummond and K. V. Kheruntsyan}
\address{{\it Department of Physics, University of Queensland, St. Lucia, Qld 4067,}\\
Australia}
\author{H. He}
\address{{\it Department of Physics, Sydney University, N.S.W 2006, Australia}}
\date{June 27, 1998}
\maketitle

\begin{abstract}
We analyze the coherent formation of molecular BEC from an atomic BEC, using
a parametric field theory approach. We point out the transition between a
quantum soliton regime, where atoms couple in a local way - to a classical
soliton domain, where a stable coupled-condensate soliton can form in three
dimensions. This gives the possibility of an intense, stable atom-laser
output.
\end{abstract}

\draft


Parametric solitons or simultaneous solitary waves (``simultons"), involving
the optical $\chi^{(2)}$ nonlinearity, have been the topic of much recent
theoretical and experimental interest in nonlinear optics. We propose a
novel mechanism by which a similar phenomenon may occur in nonlinear atomic
optics, in which coherent molecule formation in a Bose-Einstein condensate
takes the place of second harmonic generation.

This requires a coupling that converts two atoms into one molecule, thus
generating coupled atomic and molecular Bose-Einstein condensates -- and so
taking advantage of molecular states that are known to exist in alkali-metal
vapors. Our model includes a coherent molecular formation process (i.e.,
without dissipation) in an atomic BEC vapor \cite{BEC} (or atom laser \cite
{Atom laser}), either through a Feshbach resonance \cite{Feshbach} or Raman
photo-association \cite{Photoassociation}. We note that Feshbach resonances
have already been observed \cite{Feshbach-experiment}. The coherently
coupled atom-molecular condensate could provide a route to the observation
of a localized three-dimensional BEC soliton, even in the absence of a trap
potential. A possible application is in the free propagation of a
non-diverging atom laser pulse, thus greatly increasing the intensity in an
atom laser beam. Even more than this, would be the importance of observing
the striking physical properties of this novel quantum field theory, and the
corresponding Bose-enhanced chemical kinetics.

The original solution for the parametric soliton was in a one-dimensional
environment \cite{Karamzin-Sukhorukov}. These classical solutions have been
classified topologically \cite{Topological classification}, and are generic
to the mean-field theories of parametric nonlinearities that convert one
particle into two (and vice-versa). The equations are non-integrable, and
are different to the usual integrable classes of soliton equations. A
considerable advantage of these types of nonlinear equations is that they
are capable of providing solutions in one, two, or three space dimensions,
which does not occur in the usual Gross-Pitaevskii equations. Both classical 
\cite{Karamzin-Sukhorukov,Topological classification,Classical parametric
solitons} and quantum \cite{Optical mesons} solutions have been recently
identified (including observation of classical solitons in experiment \cite
{Experiments-simulton}), although these different types of soliton have
strikingly different qualitative behavior.

The purpose of this Letter is to point out the physical origin of these
differences between the quantum and classical versions of the parametric
field theory; and to identify experimental requirements for observing these
novel effects in Bose condensates. We consider the following basic
Hamiltonian, to give a simple model of molecule formation: 
\begin{equation}
\label{one}\hat H=\hat H_0+\hat H_1+\hat H_{int}~\ , 
\end{equation}
where the free and interacting Hamiltonians are: 
\begin{eqnarray}
\hat H_0&=&\hbar \int d^3{\bf x}\left[ \frac \hbar {2m}\mid \nabla \hat \Phi
\mid ^2+\frac \hbar {2M}\mid \nabla \hat \Psi \mid ^2\right] \ , 
\nonumber \\\hat H_1&=&\hbar \int d^3{\bf x}\left[ {\frac \kappa 2}\hat 
\Phi ^{\dagger 2}\hat \Phi ^2+V_\Psi ({\bf x})\hat \Psi ^{\dag }\hat \Psi
+V_\Phi ({\bf x})\hat \Phi ^{\dag }\hat \Phi \right] \nonumber \\
\hat H_{int}&=&\hbar \int d^3{\bf x}~{\frac \chi 2}\left[ \hat \Phi ^2\hat \Psi
^{\dag }+\hat \Phi ^{\dagger 2}\hat \Psi \right]  
\label{two} \ .
\end{eqnarray}
Here we define complex fields $\hat \Phi =\int d^3{\bf k}\hat a({\bf k})\exp[%
i({\bf k} \cdot {\bf x} )]$ and $\hat \Psi =\int d^3{\bf k}\hat b({\bf k})%
\exp[i({\bf k}\cdot {\bf x})]$ . The field $\hat \Phi $ represents an atomic
species of mass $m$ in a potential $V_\Phi ( {\bf x})$, in one internal
state, while $\hat \Psi $ represents a dimer species of mass $M=2m$, in a
single vibrational and rotational state, with a potential $V_\Psi ({\bf x})$.

The coupling constant $\chi $ represents a formation rate for the dimer, in
the S-wave scattering limit, while $\kappa $ represents the effective
self-interaction of the atomic field. In the absence of any trap, the
potentials are uniform, and $\hbar \rho =\hbar (V_\Psi -2V_\Phi )$ is the
formation energy of the dimer species. We note that these interactions are
idealized, in the sense that both $\chi $ and $\kappa $ represent processes
that are microscopically nonlocal. To represent such nonlocal behavior, we
must introduce a momentum cutoff $k_{m }$ in the relative momenta of
interacting fields, which physically must be around the inverse $S$-wave
scattering length -- if we wish to use the non-renormalized effective
potential to describe $S$-wave scattering. This is known to be essential to
the correct interpretation of these types of effective field theories. It
should be recognized that molecular self-interactions -- as well as
atom-molecular scattering -- will occur as well. These are neglected here,
since the relevant cross-sections are not well known.

In the corresponding nonlinear optical case, the $\Phi $ and $\Psi $ fields
would correspond to a first and second harmonic, coupled by a $\chi ^{(2)}$
nonlinearity of the dielectric, while $\kappa $ would correspond to a $\chi
^{(3)}$ nonlinearity. The interplay between quadratic and cubic
nonlinearities in the case of nonlinear optical solitons has been analyzed,
at the classical level and for one space dimension, in \cite
{Quadratic-cubic-interplay}. The effective masses, which should be different
in the longitudinal and transverse directions, describe the effects of
dispersion and diffraction, respectively, for both the fields (see, e.g., 
\cite{Optical mesons} for more details). Here the equations refer to a
moving frame situation, with coordinates moving at the group velocity.

By comparison, in the directly comparable atomic case, we are considering
atoms in free-space. No potential needs to be included, since this is not
essential to soliton formation. The molecular formation process would be
tuned in any practical experiment, by magnetic fields or external Raman
coupling, in order to reduce the energy mismatch $\hbar \rho $. An important
consideration is the possible effects of losses due to inelastic
atom-molecule collisions. We assume that an appropriate choice of molecular
levels is made, so that these losses can be ignored over the relevant
time-scales for solitons to form. Thus, the neglect of molecular vibrational
transitions is crucial to the present theory, which only includes one
molecular level. An ideal situation would involve a direct coupling via a
tuned Raman transition to the molecular ground state. A more sophisticated
theory would include detailed atomic positions and multiple energy levels
within each molecule. Our theory neglects these additional complications.

The Heisenberg equations of motion that correspond to the basic Hamiltonian
are: 
\begin{eqnarray}
i\frac \partial {\partial t}\hat \Phi &=&-\frac \hbar {2m}\nabla ^2\hat \Phi
+\chi \hat \Psi \hat \Phi ^{\dag }+\kappa \hat \Phi ^{\dagger }\hat \Phi
^2+V_\Phi ({\bf x})\hat \Phi \ , 
\nonumber \\
\label{three}i\frac \partial {\partial t}\hat \Psi &=&-\frac \hbar {2M}\nabla
^2\hat \Psi +{\frac \chi 2}\hat \Phi ^2+V_\Psi ({\bf x})\hat \Psi \ . 
\end{eqnarray}

As a first step, we can take mean values, so that $\phi =\langle \hat \Phi
\rangle $ and $\psi =\langle \hat \Psi \rangle $, and assume operator
product factorization. This gives rise to mean field equations, valid for a
momentum cutoff less than the S-wave scattering length. For the case of Bose
condensates in existing evaporative cooling experiments, near the atomic
collective ground state, the mean field equations represent modified
Gross-Pitaevskii equations -- which are known to successfully describe BEC
excitations.

Another way to understand the behavior of this quantum many-body system is
to look for energy eigenstates of the original Hamiltonian, in the limit of
a large momentum cutoff. These must simultaneously be the eigenstates of $%
\hat N=\int d^3{\bf x}\left[ |\hat \Phi |^2+2|\hat \Psi |^2\right] $,
conserving the generalized particle number $N$ (total number of atoms if we
count each molecule as two atoms). Solving this, a remarkable fact emerges.
We can show rigorously that in the limit of free space propagation, an $N$%
-boson ground state exists -- by finding exact upper and lower bounds on the
Hamiltonian energy. Since these coincide in three dimensions, we have the
result that the (idealized) quantum ground state energy is {\em exactly}:

\begin{equation}
\label{four}E_g^{N}={\frac{N }{2}}\left( \hbar \rho -{\frac{\hbar \chi ^2}{%
2\kappa }} \right) \ , 
\end{equation}
where we assume $N$ is even. The proof of the lower bound also assumes $%
\kappa >0$ and $\chi ^2>2\rho \kappa $, and the result is obtained using the
known solution of the two-particle ($N=2$) bound-state problem \cite{Optical
mesons}.

This corresponds to $N/2 $ independent quantum solitons or ``dressed''
molecules, each of which exist in a linear superposition with a pair of
atoms (like a Cooper pair), so that:

\begin{equation}
\label{five}|\psi ^{N}_Q\rangle =\left[ \hat b^{\dagger }(0)+\int_ {0}^{k_{m
}}d^3{\bf k}~g({\bf k})\hat a^{\dagger }({\bf k})\hat a ^{\dagger }(-{\bf k}%
)\right] ^{N/2} \left| 0\right\rangle \ . 
\end{equation}
In this limit of a large cutoff in the quantum field theory, the
ground-state energy has no lower bound as $\kappa \rightarrow 0$. This is in
remarkable contrast to the known mean-field behavior of the corresponding
classical energy, which is rigorously bounded below (see, e.g., \cite
{Classical parametric solitons}). Of more interest is the limiting behavior
of the ground-state quantum energy when there is a cutoff $k_{m }$ present.
We have obtained a variational estimate of this quantity, and for this case
we obtain ($\rho ,~\kappa \rightarrow 0$):

\begin{equation}
\label{six}\tilde E_Q^{N}=-{Nm\chi ^2k_{m }}/({8\pi ^2})\ . 
\end{equation}
Here we have taken the case of a relatively large cutoff, so that the result
assumes that $k_{m }\gg \left( {\chi m}/({2\pi \hbar })\right) ^2$, and uses
a variational ansatz of the form given previously. The ansatz gives us the
true ground state energy in the limit $k_{m }\rightarrow \infty $, (for any
finite $\kappa $), since upper and lower energy bounds coincide. However, it
is not necessarily the lowest possible energy at finite $k_{m }$. In order
to show this, we consider a coherent or mean-field ansatz, with broken
symmetry, of form:

\begin{equation}
\label{seven}|\psi ^{{N}}_C\rangle =\exp\left\{ \int d^3{\bf x}\left[ \phi (%
{\bf x} )\hat \Phi ^{\dagger }({\bf x})+\psi ({\bf x})\hat \Psi ^{\dagger }(%
{\bf x} )\right] \right\} \left| 0\right\rangle \ . 
\end{equation}

For this case, the classical decorrelation originates in coherent-state
factorization properties of the Hamiltonian. This state is, however, not an
eigenstate of $\hat H$ (since it is not an eigenstate of $\hat N$). It is an
approximate (semi-classical) eigenstate at large $N$, and corresponds to two
coupled Bose-Einstein condensates under broken symmetry conditions.

We will now show that, provided $\psi ({\bf x})$, $\phi ({\bf x})$ are
chosen to minimize the classical Hamiltonian, they can give a lower energy
than previously -- although still finite. This calculation makes use of the
known result that the classical parametric Hamiltonian is always bounded
below \cite{Classical parametric solitons}, and the bound is given by the
soliton energy for exact phase matching $\rho =0$. This soliton energy is
estimated by means of a variational ansatz applied to the Hamiltonian. We
choose 
\begin{eqnarray}
\phi ({\bf x})&=&g_1N^2[2/(\pi s_1)]^{3/4}\exp (-|{\bf x}|^2N^2/s_1)\ , 
\nonumber \\
\label{eight}\psi ({\bf x})&=&-g_2N^2[2/(\pi s_2)]^{3/4}\exp (-|{\bf x}|^2N^2/s_2)\ . 
\end{eqnarray}
The negative sign for $\psi ({\bf x})$ ensures that the coupling energy is
negative, and the normalization implies that $g_1^2 +g_2^2 =1$. We note that
although a uniform variational ansatz is possible, it is known that a
uniform field of this type is always unstable for a purely parametric
coupling\cite{HDM96} -- and hence cannot give the lowest energy.

Substituting into the Hamiltonian, gives us the result: 
\begin{eqnarray}
E^{N}_C/\hbar &=&N^3\left( \frac{3\hbar}{2m}\right) 
\left[ {g_1^2 \over s_1} +
 \frac {g_2^2}{2s_2} -\frac{\tilde\chi
g_1^2g_2s_2^{3/4}}{(s_1+2s_2)^{3/2}}\right] 
\nonumber \\
\label{nine}&+&\ N^5 \tilde\kappa 
 g_1^4  s_1^{-3/2}\ +\ N\rho g_2^2\ , 
\end{eqnarray}
where we have used $M=2m$, with the simplified notation of $\tilde \chi =
2^{5/2}(2/\pi)^{3/4}m\chi/(3\hbar) $ and, similarly: $\tilde \kappa =
2^{-5/2}(2/\pi)^{3/2}\kappa $.

To minimize $\tilde E_C^{N}$, under the constraint of a fixed $N$, is a
non-trivial algebraic procedure. However, the physics is considerably
simplified in the region where the term in $N^3$ is dominant -- which we
note should not involve too large a contribution from the repulsive term
that scales with $N^5$, and tends to destabilize soliton formation. In this
region (i.e., assuming $\kappa \simeq \rho \simeq 0$), we obtain a coupled
molecular Bose condensate minimum energy of: 
\begin{equation}
\label{ten}\tilde E_{C}^{N}=-CN^3\left( \frac{\hbar ^2}m\right) \left( \frac{%
m\chi }\hbar \right) ^4\ , 
\end{equation}
where $C$ is a constant given by $C\simeq 1\cdot 2\times 10^{-5}$. The
relevant length scale is nearly identical for the two coupled condensates,
and is given by: 
\begin{equation}
\label{eleven}l_1=\frac{\sqrt{s_1}}N\simeq 1.7\times 10^2~\frac 1N\left( 
\frac \hbar {m\chi }\right) ^2\ . 
\end{equation}

This enables us to more clearly understand the apparent paradox that a full
quantum theory gives a qualitatively different lower energy bound to the
corresponding classical mean field theory. To obtain a stable coupled
atom-molecular condensate, we require $\tilde E_C^N\le \tilde E_Q^N$, which
occurs at a critical boson number: 
\begin{equation}
\label{twelve}N\ge N_{cr}=\sqrt{\frac{k_m}{8\pi ^2C}}\frac \hbar {m\chi }\ .
\end{equation}

This question is therefore a subtle combination of momentum cutoff and
particle density effects. To give some numerical results we consider $m\sim
10^{-25}\,$kg, and use a $\chi $-value estimate of about $\chi \sim 10^{-6}\,
$m$^{3/2}$/sec (given in \cite{Feshbach}, by Tommasini {\it et. al.}, for a
Feshbach resonance \cite{Feshbach-experiment}), leading to $m\chi /\hbar
\simeq 10^3\,$m$^{-1/2}$. With a choice of the cutoff at $k_m\sim 1\,$nm$%
^{-1}$, this gives a critical atom number of about $N_{cr}\sim 10^3$, which
is well within the range of current BEC experiments. At low particle
density, the formation of individual dressed molecules is favored, as atoms
couple to molecules in a particle-like way. The process is analogous to Rabi
oscillations of atoms between two different electron sub-levels, except that
is occurs between pairs of atoms and the corresponding molecular levels.
These dressed states have interesting properties, reminiscent of Cooper
pairs, but cannot be described by the classical parametric soliton
equations. At large couplings $\chi $, and at large density (but not too
large so that $S$-wave scattering is dominant) the coherent coupling of two
entire condensates is dominant - just as in nonlinear optics. In this
domain, provided other recombination processes are negligible, there are
strong, coherent and nonlinear wave-like interactions between the atomic and
the molecular Bose condensates. For these parameters, it even appears
possible to form a stable, three-dimensional, Bose-Einstein soliton.

\acknowledgements

We would like to gratefully acknowledge the hospitality of the ITP
(University of California, Santa Barbara), and useful discussions with D.
Heinzen. This research was supported in part by the Australian Research
Council, and by the National Science Foundation under Grant No. PHY94-07194.

\end{document}